\documentclass[12pt]{article}
\usepackage{graphicx}
\usepackage{subfigure}
\usepackage{amsmath}
\usepackage{amssymb}

\usepackage{lipsum}
\usepackage{blindtext}

\makeatletter
\newenvironment{tablehere}
{\def\@captype{table}}
{}

\makeatother

\begin{document}
\begin{center}
\begin{large}
\title\\{ \textbf{Masses of Heavy-Light mesons with two loop static potential in a Variational approach}}\\\
\end{large}

\author\

\textbf{$ Jugal\;Lahkar^{\emph{a}}\footnotemark \:\:,Rashidul\:Hoque^{\emph{a,b}} and D.K.\:Choudhury^{\emph{a,b}}$}\\

\footnotetext{Corresponding author. e-mail :  \emph{neetju77@gmail.com}}
\textbf{a}. Dept.of Physics, Gauhati University, Guwahati-781014, India.\\
\textbf{a}. Centre of Theoretical Physics,Pandu college, Guwahati-781012, India.\\

\begin{abstract}
In this work,we investigate the masses of a few heavy-light mesons with variational method,taking into account the two-loop effects in the static Cornell potential.  Specifically,we consider two wave-function,Gaussian and Coulomb.  Our analysis suggests that phenomenologically such approach is  more successful in the Heavy flavoured meson sector than the stationary state perturbation theory and other approximation methods.

\end{abstract}
\end{center}
\

 Study of Heavy Flavour mesons are very important in heavy hadron spectroscopy as it belong to the non-perturbative aspect of QCD.  In the non-perturbative QCD potential model approach [1] have been found to be phenomenologically successful.  There are several well accepted potential available in the literature:\\
 $(1)$ Cornell potential: $ V(r)=-\frac{4\alpha_s}{3r}+br+C$[2]\\
 $(2)$ Martin potential:  $V(r)=A+Br^{\alpha}$[3]\\
 $(3)$ Logarithmic potential: $V(r)=A+Blnr$[4]\\
 $(4)$ Richardson potential: $V(r)=Ar-\frac{B}{rln{\frac{1}{\Lambda^r}}}$[5]\\
 $(5)$ Song and Lin potential: $V(r)=Ar^{\frac{1}{2}}+Br^{-\frac{1}{2}}$[6]\\
 
Among these the Coulomb plus Linear potential(Cornell) is found to be phenomenologically successful in the study of Heavy Flavour physics.  The Cornell potential shows two kinds of asymptotic behaviours: ultra-violate at short distance(Coulomb) and infra-red at long distance(Linear Confinement).  The Schrodinger equation with the Cornell potential does not provide exact analytical solutions. Although it can be solved numerically [7], [8], it is interesting to obtain analytical solutions (at least approximate ones) that offers possibility of different applications. \\
Application of variational method in Heavy quark physics was first started by Hwang etal. using linear cum coulomb potential [9], which was later successfully applied by Rai et.al [10] to find masses and decay constants of heavy flavour mesons using a power law potential($\sim \frac{-A}{r}+br^{\nu}$).  While the former[9] calculated only masses and decay constants of pseudo-scalar mesons and the later[10] also calculated the mass difference between pseudo-scalar and vector mesons . More recently, the variational method is applied to heavy flavour physics by Vega and Flores using super-symmetric potential[11]. However,one fundamental limitation of the  approach is that Heavy-Light meson can't be considered as isolated valance quark tethered to a heavy quark via a static potential.  The light quark may interact virtually with light mesons such as $\pi,k$ etc and at one loop level these interactions can explain the absence of spin orbit inversion as predicted by the potential model calculations.  We put forward a simple prescription for this by incorporating the effect of a two loop static potential using V-scheme[12].  V-scheme is a physically based alternative to the $\bar{MS}$ scheme.  Also in lattice calculations $\alpha_v$ is regarded as better expansion parameter.\\

 In this present study, we apply variational method with particularly two trial wave-functions, Gaussian and Coulomb, to study the static properties,specifically masses of pseudo-scalar heavy light mesons and mass difference between pseudo-scalar and vector mesons and then compare our results with the results of other more advanced approaches like Lattice QCD[13], QCD sum rules[14] as well as with present experimental data[15].
\\ \\

V-scheme is a popular way of taking into account the higher order effects of QCD,which are expressed as power series  in the strong coupling constant $\alpha_{\overline{MS}}$ .The two loop static potential in V-scheme is defined as[12],[16],
\begin{equation}
V(r)=-\frac{C_F\alpha_ v(\frac{1}{r^2})}{r},
\end{equation} \\
 Here,$\alpha_v$ is the effective coupling constant and in its defination all the radiative corrections are incorporated and $C_F$ is the color factor,given as,$C_F=\frac{(N_C^2-1)}{2N_C}$,where $N_C$ is the number of colors.  Generally,the quark gluon interaction is characterised by strong coupling constant $\alpha_{\overline{MS}}(q^2)$ in a dimensionally independent $\overline{MS}$ scheme[17].  V-scheme is an alternative to this.  Also in lattice calculations,$\alpha_v$ is regarded as better expansion parameter.   Recently,Schroder has improved the work [16],and the relation between $\alpha_{\overline{MS}}(q^2)$ and $\alpha_v(\frac{1}{r^2})$ at NNLO level is given by,
\begin{equation}
 \alpha_v(\frac{1}{r^2})=\frac{\alpha_{\overline{MS}}}{4\pi}(a_1+\beta_0L)+(a_2+\frac{\alpha_{\overline{MS}}^2}{16\pi^2}\beta_0(L^2+\frac{\pi^2}{3})+(\beta_1+2 \beta _0a_1)L)+O(\alpha^3_{\overline{MS}}),
\end{equation}  \\
Where,
\begin{eqnarray}
L=2ln((\mu r) (exp \gamma)),  \\
\beta_0=\frac{11}{3}C_A-\frac{4}{3}T_Fn_F ,  \\
\beta_1=\frac{34}{3}C_A^2-4C_FT_Fn_f-\frac{20}{3}C_AT_Fn_f,
\end{eqnarray}\\
And the one-loop and two loop coefficients are,\\
\begin{eqnarray}
a_1=\frac{31}{9}C_A-\frac{20}{9}T_Fn_f  ,   \\
a_2=(\frac{4343}{162}+4\pi^2-\frac{\pi^2}{4}+\frac{22}{3}\rho(3))C_A^2-(\frac{55}{3}-16\rho(3))C_FT_Fn_f+\frac{400}{81}T_F^2n_f^2-[(\frac{1798}{81}+\frac{56}{3}\rho(3))C_AT_Fn_f],
\end{eqnarray}\\\\\\

With well known QCD parameter values, $N_c=3, C_F=\frac{4}{3}, T_F=\frac{1}{2}, C_A=3,\rho(3)=1.202, \gamma_E=0.5772$.There are numerous choices for a renormalisation group improvement, i.e. for an optimal
choice of the MS scale parameter $\mu$, in order to reduce large logarithmic corrections.
There are three choices[17],\\
\pagebreak
\begin{eqnarray}
\alpha_v(\frac{1}{r^2})=\alpha_{\overline{MS}}[1+\frac{\alpha_{\overline{MS}}}{\pi}(5.76-1.03n_f)+\frac{\alpha^2_{\overline{MS}}}{\pi^2}(80.76-12.06n_f+0.3n_f^2)],
\end{eqnarray}\\

With, $\hat{\mu}=\frac{1}{r}$(made in order to eliminate r dependence in L altogether.
), \\
\begin{eqnarray}
\alpha_v(\frac{1}{r^2})=\alpha_{\overline{MS}}[1+\frac{\alpha_{\overline{MS}}}{\pi}(2.6-0.3n_f)+\frac{\alpha^2_{\overline{MS}}}{\pi^2}(53.4-7.2n_f+0.2n_f^2)] ,
\end{eqnarray}\\
 with, $\mu=\frac{exp(-\gamma_E)}{r}$(elliminates the one loop co-efficient completely),\\
 \begin{equation}
 \alpha_v(\frac{1}{r^2})=\alpha_{\overline{MS}}[1+\frac{\alpha^2_{\overline{MS}}}{\pi^2}(60.1-8.6n_f+0.2n_f^2)+(\frac{63-10n_f}{99-6n_f})(7.83+3.18n_f-0.09n_f^2)],
 \end{equation}\\
  with $\mu=\frac{exp(-\gamma_E-\frac{a_1}{2\beta_0})}{r}$(removes all terms involving the Euler constant $\gamma_E$).\\
  For $n_f=4,5$, we calculate the values of $\alpha_v(\frac{1}{r^2})$ and tabulate below :\\
 \begin{table}[h]
\caption{Values of $\alpha_v(\frac{1}{r^2})$ for different choices of $\mu$
 :}
\label{Table5}
\begin{center}
\begin{tabular}{|l|c|c|c|}
\hline
 Choices&$\mu=\frac{1}{r}$& $\mu=\frac{exp(-\gamma_E)}{r}$ & $\mu=\frac{exp(-\gamma_E-\frac{a_1}{2\beta_0})}{r}$ \\ \hline
$\alpha_{\overline{MS}}=0.22,n_f=5$ &0.26    & 0.261   &  0.256             \\ \hline
$\alpha_{\overline{MS}}=0.39,n_f=4$ & 0.69& 0.65     &  0.60                      \\ \hline

\end{tabular}
\end{center}
\end{table}

 Variational method is one of the oldest and popular approximation methods of Quantum mechanics.  The ultimate success of variational method depends upon the choice of proper wave-function. Let us consider the Gaussian wave-function as the trial wave-function :\\

\begin{equation}
\psi(r)={(\frac{\mu \alpha}{\sqrt{\pi}})^{\frac{3}{2}}}e^{-\frac{\alpha^2 r^2}{2}},
\end{equation}\\
where, $\alpha$ is the variational parameter.  We choose the  $Q\bar{Q}$ potential to be two loop modified Cornell potential as,\\
\begin{equation}
V(r)=-\frac{C_F\alpha_ v(\frac{1}{r^2})}{r}+br,
\end{equation}\\
where, $C_F=\frac{4}{3}$ is the color factor and $b$ is the standard confinement parameter.  The ground state energy is given by,\\

\begin{equation}
E(\alpha)=\langle\psi\vert H \vert\psi\rangle ,
\end{equation}
\pagebreak
With the Gaussian wave-function we obtain finally,\\
\begin{equation}
E(\alpha)=-\frac{3\mu^2 \alpha^2}{4}-\frac{8\mu^3 \alpha \alpha_v(\frac{1}{r^2})}{3\sqrt{\pi}}+\frac{2\mu^3 b}{\sqrt{\pi}\alpha},
\end{equation}\\

 Now,by minimising $E(\alpha)$ with respect to $\alpha$ we can find the variational parameter $\alpha$ for different heavy flavoured mesons.  The minimization condition to find the expectation value of Hamiltonian as,
\begin{equation}
\frac{dE(\alpha)}{d\alpha}=0 ,
\end{equation}\\
at $\alpha={\alpha'}$.  Now,putting equation $(15)$ in $(16)$, we get, \\
\begin{equation}
\frac{3}{2}\mu^2  \alpha'^3 +\frac{8\mu^3  \alpha_v(\frac{1}{r^2})\alpha'^2}{\sqrt{\pi}}-\frac{\mu^3 b}{\sqrt{\pi}}=0 ,
\end{equation}\\

This equation is solved by using $Mathematica7$ and we find the variational parameter for different Heavy Flavour mesons which is shown in $Table.2$.

\begin{table}[h]
\caption{Values of variational parameter for different heavy-light mesons for $\alpha_v(\frac{1}{r^2})$ for $\mu\sim \frac{1}{r}$
 :}
\label{Table5}
\begin{center}
\begin{tabular}{|l|c|}
\hline
 Mesons &variational parameter $(\alpha')$   \\ \hline
 $D{(c\overline{u}/\overline{c}d)}$&0.531             \\\hline
$D{(c\overline{s})}$&0.579                      \\\hline

$B{(u\overline{b}/d\overline{b})}$ & 0.593 \\\hline
    $B_s{(s\overline{b})}$&0.655 \\\hline
    $B{(\overline{b}c)}$ &0.869 \\\hline

\end{tabular}
\end{center}
\end{table}

For a second analysis, we consider a Coulombic trial wave-function as,\\
\begin{equation}
\psi(r)=\frac{(\mu \alpha')^{\frac{3}{2}}}{\sqrt{\pi}}e^{-\mu \alpha' r},
\end{equation}\\
Here,as well as in equation(11) $\mu=\frac{m_Qm_{\bar{Q}}}{m_Q+m_{\bar{Q}}}$ ,is the reduced mass. Now,with the potential as defined in equation $(12)$, following the same procedure as in the previous section, we get,\\
\begin{equation}
E(\alpha')=<\psi\mid H\mid \psi> = \frac{1}{2}\mu \alpha'^2-A\mu \alpha'+\frac{3b}{2\mu \alpha'},
\end{equation}\\
where, $A=\frac{4\alpha_v(\frac{1}{r^2})}{3}$.  Now, minimising, $\frac{dE}{d\alpha'}=0$ , we get,\\
\begin{eqnarray}
\alpha'^3-A\alpha'^2-\frac{3b}{2\mu^2}=0,
\end{eqnarray}\\
This equation is solved by using $Mathematica7$ and we find the variational parameter for different Heavy Flavour mesons which is shown in $Table.3$.\\

\begin{table}[h]
\caption{Values of variational parameter for different heavy-light mesons
 :}
\label{Table5}
\begin{center}
\begin{tabular}{|l|c|}
\hline
 Mesons &variational parameter $(\alpha')$   \\ \hline
 $D{(c\overline{u}/\overline{c}d)}$&1.72868             \\\hline
$D{(c\overline{s})}$& 1.466     \\\hline

$B{(u\overline{b}/d\overline{b})}$ & 1.531\\\hline
    $B_s{(s\overline{b})}$& 1.250\\\hline
    $B{(\overline{b}c)}$ &0.722\\\hline

\end{tabular}
\end{center}
\end{table}

Pseudo-scalar meson mass can be computed from the following relation [18], [19] :\\
\begin{eqnarray}
M_P = M+m+\frac{p^2}{2m}+ \triangle E  ,
\end{eqnarray}\\

Here, $M$ and $m$ are the masses of Heavy quark/anti-quark and light quark/anti-quark respectively and we have considered 1st order relativistic correction to the light quark/anti-quark. The energy shift of mass splitting due to spin interaction in the perturbation
theory reads[19], [20], [21], [22],\\
\begin{eqnarray}
\triangle E=\int \psi^{*}(\frac{32\pi\alpha_s}{9}\delta^3 (r)\frac{\bold{S_Q.S_{\bar{Q}}}}{M.m})\psi d^3r,
\end{eqnarray}\\
leading to,\\
\begin{equation}
\triangle E=\frac{32\pi\alpha_s}{9Mm}{\bold{S_Q.S_{\bar{Q}}}}{\mid\psi(0)\mid}^2,
\end{equation}\\
For pseudo-scalar mesons, $\bold{S_Q.S_{\bar{Q}}}=-\frac{3}{4}$, so pseudo-scalar meson masses can be expressed as,\\
\begin{equation}
M_P = M+m+\langle -\frac{\nabla^2}{2m}\rangle-\frac{8\pi\alpha_s}{3Mm}{\mid\psi(0)\mid}^2,
\end{equation}\\
This is to be contrasted with some of the previous communications [23, 24, 25], where spin dependence and spherically symmetric property of the $Q\bar{Q}$ system was not taken into account for $l=0$ state.
The mass difference between the Pseudo-scalar and the vector meson is given by[26],\\
\begin{equation}
M_v -M_p=\frac{8\pi A}{3m_Qm_{\bar{Q}}}{\mid\psi_{Q\bar{Q}}(0)\mid}^2,
\end{equation}\\
where $m_Q$ is the mass of heavy quark and $m_{\bar{Q}}$ is the mass of antiquark.
This is attributed to the hyperfine interaction and $A=\frac{4\alpha_v}{3}$,  where $\alpha_v$ is the strong coupling constant in two loop static potential. \\\\

With the formalism developed in the previous sections, we calculate the masses  of some  Heavy-Light mesons, which are shown in $Table.4$ . The input parameters are $m_{u/d}=0.336Gev$, $m_b=4.95GeV$, $m_c=1.55GeV$, $m_s=0.483GeV$ and $b=0.183GeV^2$[15][26], also we take $\alpha_s=0.39$ for c-scale and $\alpha_s=0.22$ for b-scale [26]. We  make a comprehensive comparison of our results with lattice results[13], QCD sum rule[14], previous model based on perturbation theory [23] and present experimental results [15]. Our result agrees well with the present results of  lattice QCD[13], QCD sum rules[14] and experimental data[15], than the previous models which were based on Dalgarno's perturbation theory[23],[24].  As an illustration, from Table.4, the predicted  mass of D and B meson with Gaussian trial wave-function are ($1.87$GeV and $5.27 $GeV), which are quite close to lattice results ($1.885$GeV and $5.283$ GeV)and QCD sum rule results ($1.87$ GeV and $5.28$GeV).  The pattern is similar to other heavy-light mesons. Also,the mass difference of pseudo-scalar and vector mesons obtained with Gaussian trial wave-functions agrees with present lattice results[27](Table.5).

\begin{tablehere}\scriptsize
\begin{center}
\caption{Masses of Heavy Flavoured mesons(in GeV)}
\begin{tabular}{|c|c|c|c|c|c|c|}
  \hline
          
    Mesons  &  $M_{P}$(Gaussian) & $M_{P}$(Coulomb) &  [23] & lattice[13]&Q.sum rule[14]& Exp.Mass [15]  \\
    \hline
    $D{(c\overline{u}/cd)}$   &1.878 &1.606& 2.378 &1.885&1.87  & $1.869\pm0.0016$  \\ \hline
    $D{(c\overline{s})}$  & 2.01&1.739 &2.5&1.969&1.97& $1.968\pm0.0033$  \\ \hline
    $B{(u\overline{b}/d\overline{b})}$   &5.28&5.11 & 5.798&5.283 &5.28&$5.279\pm0.0017$   \\\hline
    $B_s{(s\overline{b})}$&  5.4 &5.40 &5.331&5.366&5.37&$5.366\pm0.0024$     \\\hline
    $B{(\overline{b}c)}$ & 6.48&6.38&6.5 &6.278  && $6.277\pm0.006$  \\
    \hline
\end{tabular}
\end{center}
\end{tablehere}

\begin{tablehere}\scriptsize
\begin{center}
\caption{Mass difference of pseudoscalar and vector mesons(in GeV):}
\begin{tabular}{|c|c|c|c|}
   \hline
    Mesons&$M_V-M_P$(Gaussian)& $M_V-M_P$(Coulomb) &$M_V-M_P$(Lattice) [27]        \\\hline  
    $D{(c\overline{u}/cd)}$&0.08&0.44&      0.067                    \\\hline  
   $D{(c\overline{s})}$ &0.025&0.51&       0.066                       \\\hline
   $B{(u\overline{b}/d\overline{b})}$&0.002&0.04&   0.034              \\\hline
   $B_s{(s\overline{b})}$&0.01&0.06&      0.027                    \\\hline
   $B{(\overline{b}c)}$ &0.07&0.062&                       \\\hline
\end{tabular}
\end{center}
\end{tablehere}

        In this work we have reinvestigated the masses  of a few heavy-light mesons using variational scheme with introduction of a two loop static potential. The present analysis is done to provide a simple prescription to apply variational method in heavy-light systems. Because, Heavy-Light meson can't be considered as isolated valance quark tethered to a heavy quark via a static potential.  The light quark interact virtually with light mesons such as $\pi,K$ etc and at one loop level these interactions can explain the absence of spin orbit inversion as predicted by the potential model calculation. Therefore,the introduction of two loop static potential is necessary.\\

    The results agrees well with those obtained from more advanced approaches like lattice QCD [13] and QCD sum rules [14] and also with recent experimental data[15].  Phenomenologically, variational method with two loop static potential with Gaussian wave function provides a simple insight into the static properties of heavy-light mesons. This encourages for further application of the approach to investigate static and dynamic properties of Heavy-Light systems.
    One fundamental limitation is that,application of non-relativistic approach to heavy-light mesons is not fully justified, as one quark is indeed light and relativistic. However, the model can provide good estimate of the static  properties of Heavy-Light mesons, at least,at the phenomenological level.  To conclude,the variational method with a Gaussian trial wave-function provides a simple method to study the static and dynamic properties of pseudo-scalar mesons which are close to the corresponding results of the lattice QCD and QCD sum rules.  Such effective H.O. wave function can presumably be  generated by a  $Q\bar{Q}$ potential, which is polynomial in r,  $V(r)=\Sigma_{n=-l}^{n=+l}a_nr^n, with, a_2 \gg a_{l,l\neq2}$,  a feature noticed by Godfrey and Isgur[28].\\\\
  \paragraph{Acknowledgement:}
\begin{flushleft}
\emph{One of the authors (Jugal Lahkar) acknowledges the financial support of CSIR(New-Delhi,India) in terms of fellowship under Net-Jrf scheme to pursue research work at Gauhati
University, Department of Physics. We also thank Dr.B.J.Hazarika of Pandu College,Ghy-781012,India and Dr.N.S.Bordoloi of Cotton University,Ghy-781001,India for various useful discussions.}
\end{flushleft}

\end{document}